\documentclass[twocolumn]{aastex631}

\usepackage{booktabs} 
\usepackage{graphicx}  

\usepackage{soul}

\newcommand{\scs}{\textsc}

\begin{document}



\title{Discovery of a barium blue straggler star in M67 and `sighting' of its WD companion\footnote{This is paper XV of the UVIT Open Cluster Study (UOCS)}}

\author[0009-0001-8079-3471]{Harshit Pal}
\affiliation{Indian Institute of Astrophysics, Koramangala II Block, Bangalore, 560034, India}
\affiliation{Indian Institute of Science Education and Research, Berhampur, Odisha, 760010, India}

\author[0000-0003-4612-620X]{Annapurni Subramaniam}
\affiliation{Indian Institute of Astrophysics, Koramangala II Block, Bangalore, 560034, India}

\author[0000-0003-2754-641X]{Arumalla B. S. Reddy}
\affiliation{Indian Institute of Astrophysics, Koramangala II Block, Bangalore, 560034, India}

\author[0000-0002-8672-3300]{Vikrant V. Jadhav}
\affiliation{Helmholtz-Institut für Strahlen- und Kernphysik, Universität Bonn, Nussallee 14-16, D-53115 Bonn, Germany}

\begin{abstract}

We report the discovery of a barium blue straggler star (BSS) in M67, exhibiting enhancements in slow neutron-capture ($s$-) process elements. Spectroscopic analysis of two BSSs (WOCS\,9005 \& WOCS\,1020) and 4 stars located near the main-sequence turn-off using GALAH spectra, showed that WOCS\,9005 has a significantly high abundance of the s-process elements ([Ba/Fe] = 0.75$\pm$0.08, [Y/Fe] = 1.09$\pm$0.07, [La/Fe] = 0.65$\pm$0.06).  The BSS (WOCS\,9005) is a spectroscopic binary with a known period, eccentricity and a suspected white dwarf (WD) companion with a kinematic mass of 0.5 M$_\odot$. The first `sighting' of the WD in this barium BSS is achieved through multi-wavelength spectral energy distribution (SED) with the crucial far-UV data from the UVIT/\textit{AstroSat}. The parameters of the hot and cool companions are derived using binary fits of the SED using two combinations of models, yielding a WD with  T$_{eff}$ in the range 9750--15250 K. 
Considering the kinematic mass limit, the cooling age of the WD is estimated as $\sim$ 60 Myr. The observed enhancements are attributed to a mass transfer (MT) from a companion asymptotic giant branch star, now a WD. We estimate the accreted mass to be 0.15 M$_{\odot}$, through wind accretion, which increased the envelope mass from 0.45 M$_{\odot}$. The detection of chemical enhancement, as well as the sighting of WD in this system have been possible due to the recent MT in this binary, as suggested by the young WD.


\end{abstract}


\section{Introduction} \label{sec:intro}

Star clusters are test beds to understand single and binary star evolution. Clusters host a good fraction of binaries \citep{Jadhav2021AJ....162..264J} as well as binary products due to evolution and stellar encounters. Among the binaries, the post mass-transfer (MT) binary systems play an important role in the formation of Type Ia supernovae, blue straggler stars (BSSs), sub-dwarfs and extremely low mass (ELM) white dwarfs (WDs). 

The chemical enhancement in the secondary companion due to MT from an evolved primary in a binary has been a topic of research and in particular, transfer of $s$-process elements synthesized during the thermally pulsing phase of asymptotic giant branch (AGB) companion \citep{busso1999}. Even though binaries are plentiful in clusters, only a handful of chemically enriched post-MT binaries have been found \citep{martins2017properties, van2017mass, li2022discovery}. The BSSs are likely to be formed via MT \citep{mccrea1964extended} or mergers \citep{andronov2006mergers} and those formed through a MT in which the primary evolved through an AGB phase (Case C MT \citealt{chen2008mass}), are potential targets for chemically enriched systems.
There is one confirmed Barium star (a giant) known in NGC\,2420 \citep{van2017mass}, and its orbital properties confirm the presence of an unseen WD. The cluster NGC 5822 has two s-process enhanced giants \citep{santrich2013two}. The cluster NGC\,6819 has a double-lined spectroscopic binary BSS with Barium enhancement and 4 barium enhanced BSSs with no radial velocity variations \citep{milliman2015barium}. \citet{mathys1991blue} detected two Ba-enriched BSSs in M67 (S968 and S1263), which are found to be single members \citep{geller2015}.
This is indeed puzzling as there are many BSSs that are single lined spectroscopic binaries (SB1s) with unseen WD binaries and should have gone through a Case C MT with detectable chemical enrichment. 

In the Galactic field, barium and other $s$-process elements are found to be enhanced in  barium (Ba {\scs II}) stars \citep{mcclure1984barium}. These stars are widely studied \citep{liu2009abundance}, where many are giants (giant barium stars) \citealt{jorissen2019barium} and a few are in the main-sequence (MS) (dwarf Barium stars) \citep{escorza2019barium}. Since \citet{mcclure1980binary} and \citet{mcclure1984barium} showed that Ba {\scs II} stars are in binaries, it is widely accepted that their enhanced elements came from a companion AGB star, which is now an optically invisible/unseen WD. 
Therefore, there is a sharp contrast as 
 only a few enhanced systems are seen in the BSSs in star clusters that are formed via MT process.

\citet{gray2011first} found that barium stars have a real far-UV excess based on the analysis of 6 Barium dwarfs and F-type dwarfs. 
Secondary stars of a few planetary nebulae have been found to be barium enhanced, and many studies have characterised the WD in these systems using UV data (\citealt{siegel2012swift}  \citealt{miszalski2012barium, miszalski2013}, \citealt{merle2016ba}).



M67 (Messier 67; NGC\,2682), an old and rich open cluster with solar-like age and metallicity \citep{hobbs1991metallicity}, is a goldmine to study stellar evolution, stellar dynamics and binary products. Most notably, this cluster houses a large number of BSSs \citep{liu2008spectroscopic, Jadhav2021MNRAS.507.1699J} and the radial velocity surveys have shown a much higher binary fraction among BSSs, 60 ± 24 percent \citep{Latham2007, geller2008}.
There have been attempts to search for chemical enhancements among the BSSs of M67 \citep{mathys1991blue, bertelli2018chemical, brady2023m67}.
With the help of UV photometric data, several studies reported the detection of ELM WD companions to some of the BSSs \citep{uocsi, uocsii,pandey2021uocs}. Radial velocity monitoring studies of M67 have been extensively carried out under the WIYN open cluster study (WOCS) program, and the orbital parameters of a large number of stars are presented in \citet{geller2015} and \citet{geller2021}.

WOCS\,9005 is a BSS that is a single lined spectroscopic binary (SB1), but located close to the main-sequence turn-off (MSTO) \citep{geller2015}. \citet{uocsii} detected this star in one filter in the FUV and found a possible hot-companion but mentioned that further observations are needed to confirm the presence of the hot companion. \cite{leiner2019} estimated a probable rotation of 4.5 days, estimated the unseen WD mass to be $\sim$ 0.5 M$_\odot$ and estimated a gyro-age of $\sim$ 500 Myr. \citep{geller2021} published the orbital parameters as well as radial velocity curves.
 
Here, we present the discovery of WOCS\,9005 as a barium BSS with s-process enhancement based on s-process elemental abundances from GALAH spectra \citep{galah}. We also present the first `sighting' of the WD in the form of excess far-UV flux and estimate the WD parameters from the fits to its spectral energy distribution (SED). This work has many implications, such as the  confirmation of the MT process by `seeing the unseen WD' in a barium BSS, binary MT process at work in the cluster environment, and the physics of accretion processes through stellar wind-accretion in a binary system. 


\section{Data sets} \label{sec:data}

We have used the UVIT (Ultra-Violet Imaging Telescope (UVIT) onboard \textit{AstroSat} catalog of M67 from \citet{uocsiii}, with magnitudes along with their errors in the 3 FUV filters, F148W, F154W and F169M.
We further utilized pre-existing photometric data from \textit{Swift} UVOT Open Clusters Catalog 
\footnote{\url{https://archive.stsci.edu/hlsp/uvot-oc}}
M67 in 3 UV filters (UVW1, UVM2, and UVW2), and in 1 optical filter (U; \citealt{siegel3}) and additional archival data (Section \ref{sec:SED}).

We have used the radial velocity time series data and the orbital parameters from \cite{geller2021}. GALAH is an optical spectroscopic survey that utilizes the HERMES spectrograph (R=28000) at the Anglo-Australian Telescope, offering 678,423 spectra for 588,571 stars, primarily within 2 kpc. 
We have used the abundance data for M67 stars as well as the spectroscopic data from the GALAH survey.

\begin{figure}[ht]
    \centering 
     \includegraphics[width=0.45\textwidth]{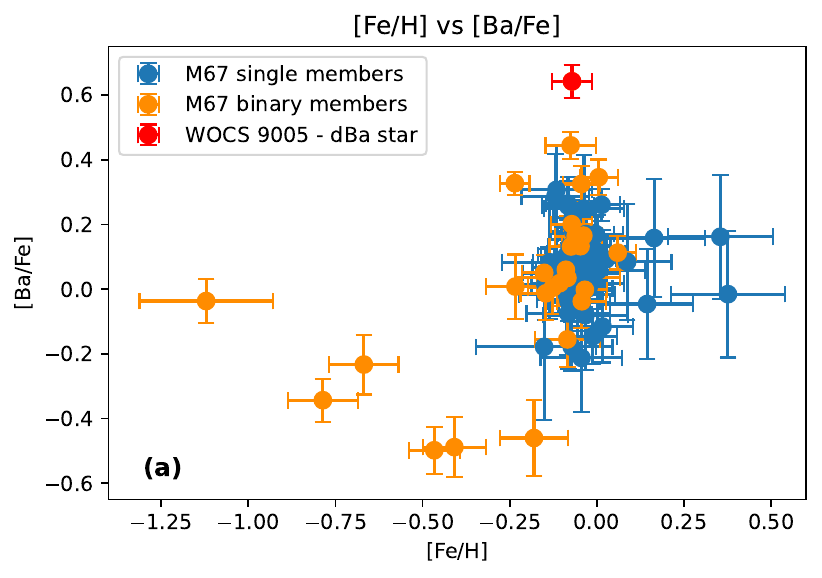}
        \includegraphics[width=0.45\textwidth]{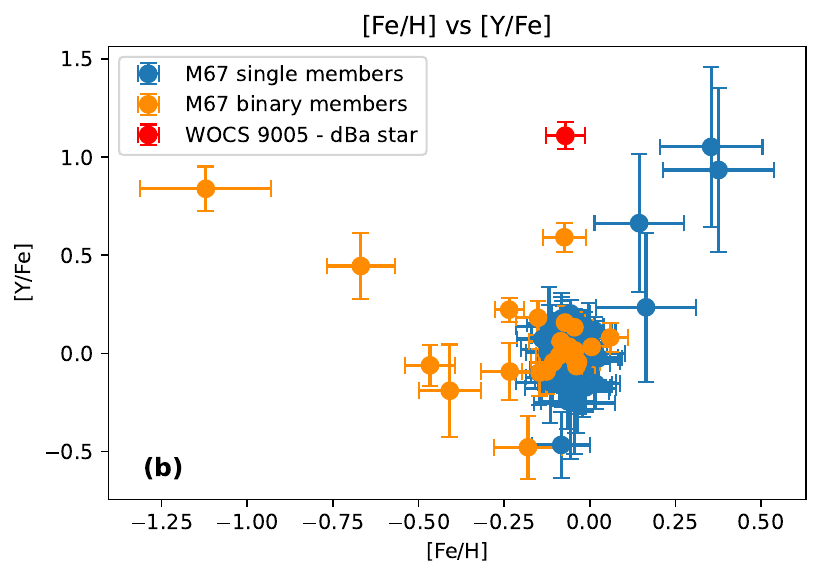}
        \includegraphics[width=0.45\textwidth]{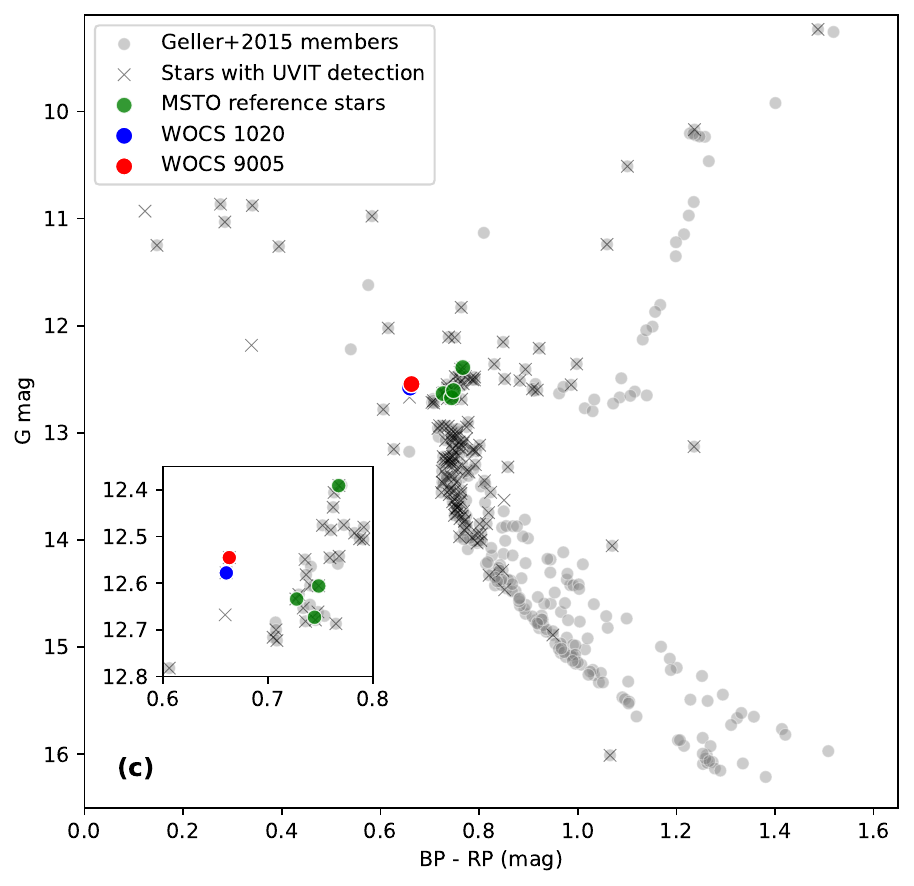}
    \caption{(a) Barium and (b) Yttrium abundances of single members (SM) and binary members (BM) of M67 (from \citealt{geller2021}) as estimated by GALAH \citep{galah} are shown.  Panel (c) shows the Color Magnitude Diagram of M67 using the Gaia DR3 data. }
    \label{fig:galah}
\end{figure}

\section{Selection of the target based on abundances from GALAH} \label{sec:target}

Numerous heavy elements were detected in the star by the GALAH survey \citep{galah}, including yttrium (Y), which is part of the first s-process peak, as well as barium (Ba) and lanthanum (La), which belong to the second s-process peak. 
We selected stars that are (1) detected in the UVIT images with a good SNR, and (2) membership from \citet{geller2021} and (3) have [Ba/Fe] and [Y/Fe] abundance estimates from GALAH. Fig \ref{fig:galah} (a-b) show the [Ba/Fe] and [Y/Fe] abundances against the [Fe/H] values of 88 stars that satisfy all of the above three conditions. Among these 88 stars, 22 have orbital parameters information in \citet{geller2021} catalog. 

From Fig \ref{fig:galah} (a), we find that a few stars have overabundance of barium with respect the majority of stars.  Among these, WOCS\,9005 is an SB1 and a BSS, which shows high barium abundance. As shown in the panel (b), the same star also has high yttrium abundance and stands out with statistically significant excess in comparison to the majority of stars. 
Therefore, WOCS\,9005 is a very strong candidate for a dwarf barium star. It has also been classified as a BSS, and we confirm the classification based on its location as shown in the Gaia colour-magnitude diagram (CMD) in Fig \ref{fig:galah} (c). This star therefore gains importance as a candidate to detect s-process chemical enhancement in a BSS, that is likely to be formed via MT.

\section{Stellar parameters and chemical composition of WOCS\,9005} \label{sec:spectroscopy}


 This section presents an independent chemical abundance estimates of the GALAH optical spectra of WOCS\,9005, and compare them with those of a similar BSS (WOCS\,1020) and 4 stars located on the MSTO of the cluster (WOCS\,4021, 5009, 4004, and 7010  marked in panel (c) of Fig \ref{fig:galah}). The chemical composition of these MSTO stars is assumed to be representative of the mean cluster abundance. The reduced one-dimensional spectra of the above stars in M67, observed at a spectral resolution of 28,000 are retrieved from the GALAH DR3 Spectral Library\footnote{\url{https://www.galah-survey.org/dr3/the_spectra/}}. The typical signal-to-noise ratio of the spectra is higher than 100 over the wavelength range covered by the high-resolution multi-fibre spectrograph, HERMES.


A differential abundance analysis relative to the Sun was performed by running the {\it abfind} driver of {\scs \bf MOOG} \citep{sneden1973phd} adopting the ATLAS9 one-dimensional, line-blanketed, plane-parallel local thermodynamic equilibrium (LTE) model photospheres computed with updated opacity distribution functions \citep{caskurucz2003} and the measured iron line equivalent widths (EWs) from the observed spectra following the standard LTE abundance analysis technique described in \citet{reddy2019comprehensive}. The stellar parameters ( effective temperature (T$_{\rm eff}$), surface gravity (log~$g$), microturbulence velocity ($\xi_{t}$) and metallicity ([Fe/H])) were derived by force-fitting the model-generated iron line EWs to the observed ones by imposing the excitation and ionization equilibrium and the requirement that the derived iron line abundances are independent of measured line EWs.

The abundances for Na and the $s$-process elements were derived by fitting synthetic profiles to the observed lines of Na (5688 \AA), Y {\scs II} (4884 and 5729 \AA), Ba {\scs II} (5854 and 6497 \AA) and La {\scs II} (4804 and 5806 \AA), following the standard synthetic profile fitting procedure \citep{reddy2019comprehensive} where the computed synthetic profiles were smoothed before matching the observed spectra with Gaussian profiles representing the instrumental and rotational ($v$\,sin$i$) broadening of the spectral lines. The line list employed for the spectrum synthesis of Ba {\scs II} lines includes hyperfine structure components and isotopic shifts \citep{mcwilliam1998}, and the fractional contribution of each isotope to solar system Ba abundances \citep{lodders2003}. 
Excluding Ti, Cr, Y {\scs II} (5729 \AA) and two La {\scs II} lines, the spectral lines used for chemical composition analysis are stronger than 10 m\AA, while the former set of lines have EW\,$<$\,10 m\AA\ and their measured abundances are sensitive to spectral noise and continuum placement. 
However, all the key lines of interest (Na, Y, Ba, and La) in the $s$-process enhanced star WOCS\,9005 are stronger than 10 m\AA\ to derive the best estimates of elemental abundances.

The final stellar parameters and derived abundances along with typical 1-$\sigma$ errors for WOCS\,9005 and the comparison star WOCS\,1020 selected based on the similarity of stellar parameters with WOCS\,9005 are provided in Table~\ref{tab:abundances}. The 1-$\sigma$ error is estimated by adding in quadrature the sum of the errors introduced by uncertainties in stellar parmeters and spectrum synthesis/EW analysis. Average [X/Fe] values for 4 turn-off stars and WOCS\,1020, are also presented in the table~\ref{tab:abundances}. The individual abundances and stellar parameters of the 4 MSTO stars are provided in Table~\ref{tab:MSTO_abundances} in Appendix. 

\begin{figure}
    \centering
    \includegraphics[trim=1.5cm 10.4cm 9.1cm 7.0cm, clip=true, width=0.47\textwidth]{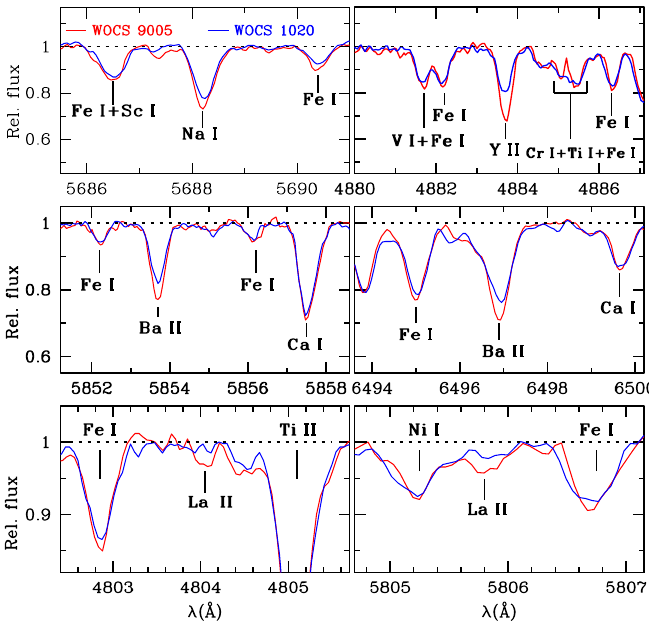}
    \caption{Observed spectra of WOCS\,9005 (red) and WOCS\,1020 (blue) covering the wavelength regions containing Na and the $s$-process elements Y, Ba, and La. The spectral lines due to other species are also shown.
    \label{fig:spectra}}
\end{figure}

Entries in Table~\ref{tab:abundances} offer the striking evidence that WOCS\,9005 is enhanced with Na and $s$-process elemental abundances relative to MSTO stars including WOCS\,1020 (i.e. within the run from Na to La, [X/Fe] ratios for WOCS\,1020 and MSTO stars are nearly solar (Table~\ref{tab:MSTO_abundances}), whereas sodium and the $s$-process elements (only) are overabundant in WOCS\,9005 over comparison sample with closely matched [X/Fe] values for all other elements).
These results agree well within $\pm$0.1 dex for almost all elements in common with the abundances reported from the GALAH survey. 
Fig.~\ref{fig:spectra} offers a direct comparison of observed spectra of WOCS\,9005 with WOCS\,1020 covering the spectral lines due to Na, and $s$-process elements Y, Ba, and La. 
The line strengths for the $s$-process elements Y, Ba and La in WOCS\,9005 are more pronounced, signifying the reliability of derived overabundances for WOCS\,9005.
 
The presence of sodium (generally observed in stars evolved to RGB/AGB phase) along with the $s$-process enrichment, binarity of WOCS\,9005 and its current location near the main sequence turn-off precludes the possibility of formation of WOCS\,9005 from the parent cluster gas cloud already enriched with Na and $s$-process elements and indicate the presence of accreted material in the photosphere of WOCS\,9005 from its evolved binary companion, which is now a WD.  \cite{leiner2019} derived a kinematic mass of $\sim$ 0.5\,$\mathrm{M}_{\odot}$ for the WD companion of WOCS\,9005 from the relation connecting the binary mass function f(m) with the radial velocity semi-amplitude, orbital period, and eccentricity. The mass of WOCS\,9005 was adopted by ﬁtting theoretical PARSEC stellar evolutionary tracks \citep{Bressan2012} to it's position in the CMD. The orbital parameters adopted by \cite{leiner2019} for WOCS\,9005 binary system are same as those presented in \cite{geller2021} via an extensive radial-velocity observations of M67 members. 
The orbital parameters and mass estimates of the binary system (WOCS\,9005 and its WD companion) are presented in table~\ref{tab:abundances}.


\begin{table*}
\centering
\caption{Elemental abundances and stellar parameters for WOCS\,9005 and WOCS\,1020 and the average [X/Fe] values for 4 turn-off stars and WOCS\,1020.  The numbers in the parentheses are the number of lines employed for the abundance measurement of an element. The estimated parameters of WOCS\,9005 and WOCS\,1020 using spectroscopy and photometry are shown below, followed by orbital parameters and mass of the components of WOCS\,9005 \citep{geller2021, leiner2019} and estimated parameters of the cooler and hotter companions from the SED fits.}
\label{tab:abundances} 
\begin{tabular}{cccc}
\toprule
Parameter             &   WOCS\,9005      &    WOCS\,1020       & Average\_{TO}   \\   \midrule
$[$Na \scs{I}/Fe$]$ & $+0.28\pm0.05(1)$   & $\,0.00\pm0.05(1)$   &  $+0.08\pm0.07$  \\
$[$Mg \scs{I}/Fe$]$ & $+0.07\pm0.04(1)$   & $+0.03\pm0.05(2)$   &  $+0.05\pm0.06$  \\
$[$Al \scs{I}/Fe$]$ & $+0.07\pm0.05(4)$   & $+0.09\pm0.03(3)$   &  $+0.08\pm0.02$  \\
$[$Si \scs{I}/Fe$]$ & $-0.04\pm0.04(4)$   & $-0.02\pm0.02(3)$   &  $\,0.00\pm0.06$ \\
$[$Ca \scs{I}/Fe$]$ & $-0.01\pm0.06(4)$   & $-0.01\pm0.05(3)$   &  $+0.03\pm0.02$  \\
$[$Ti \scs{I}/Fe$]$ & $-0.01\pm0.05(3)$   & $+0.02\pm0.04(2)$   &  $\,0.00\pm0.03$ \\
$[$Cr \scs{I}/Fe$]$ & $-0.05\pm0.03(1)$   & $-0.02\pm0.03(1)$   &  $-0.02\pm0.03$  \\
$[$Fe \scs{I}/H$]$  & $-0.10\pm0.05(21)$  & $-0.11\pm0.06(18)$  &  $-0.14\pm0.02$  \\
$[$Fe \scs{II}/H$]$ & $-0.11\pm0.07(3)$   & $-0.09\pm0.07(2)$   &  $-0.14\pm0.02$ \\
$[$Ni \scs{I}/Fe$]$ & $-0.04\pm0.05(9)$   & $\,0.00\pm0.05(7)$  &  $\,0.00\pm0.04$ \\
$[$Y \scs{II}/Fe$]$ & $+1.09\pm0.07(2)$   & $+0.05\pm0.08(2)$   &  $+0.08\pm0.02$  \\
$[$Ba \scs{II}/Fe$]$& $+0.75\pm0.08(2)$   & $+0.12\pm0.09(2)$   &  $+0.14\pm0.04$ \\
$[$La \scs{II}/Fe$]$& $+0.65\pm0.06(2)$   & $+0.09\pm0.06(2)$   &  $+0.05\pm0.07$ \\
\midrule
T$_{\text{eff}}$ (K)$^{\dagger}$    & $6350 \pm 50$  & $6350 \pm 50$    & \\
log~$g$ (cm s$^{-2}$)$^{\dagger}$   & $4.2 \pm 0.2$  & $4.2 \pm 0.2$    & \\
$\xi_{\text{t}}$ (km s$^{-1}$)$^{\dagger}$ & $1.47 \pm 0.1$ & $1.70 \pm 0.1$ &  \\
$[$Fe/H$]$ (dex)$^{\dagger}$        &  $-0.10 \pm 0.05$  & $-0.11 \pm 0.06$  &  \\
$v \sin i$ (km s$^{-1}$)$^{\dagger}$ & $15.0 \pm 2.0$    & $14.0 \pm 2.0$    & \\
$M\ (M_{\odot}$)$^{\ddagger}$ & $1.37 \pm 0.04$ & $1.30 \pm 0.05$ & \\
$R\ (R_{\odot}$)$^{\ddagger}$ & $1.99 \pm 0.06$ & $1.80 \pm 0.05$ & \\

\end{tabular}
\begin{tabular}{cccccccc}   
\toprule
  RV$_{\text{COM}}$ & Period & RV$_{semi}$ & e & $f(m)$  & a\,sini  & $\mathrm{M}_{1}$ & $\mathrm{M}_{WD}$  \\
 (km s$^{-1}$)     & (days) &(km s$^{-1}$) &   & ($10^{-2} \mathrm{M}_{\odot}$) &(10$^{11}$ m) & ($\mathrm{M}_{\odot}$)  & ($\mathrm{M}_{\odot}$)  \\ \midrule
 32.91$\pm$0.16 & 2769$\pm$18 & 5.1$\pm$0.3 & 0.15$\pm$0.05 & 3.7$\pm$0.6 & 1.92$\pm$0.11 & $1.37\pm0.04$ & $0.5$  \\
\end{tabular}
\begin{tabular}{lrrrrrr}

\toprule
Component & $T_{\text{eff}}$ $(K)$ & $R(R_{\odot}$) & $L(L_{\odot}$) & $\log g$ & $\chi_{r}^2$  & vgf$_b^2$ \\ [0.4em] \midrule 
\multicolumn{7}{c}{Kurucz + Koester} \\[0.4em]
WOCS 9005A & $6500^{+250}_{-250}$ & $1.85^{+0.03}_{-0.03}$ & $5.52^{+1.15}_{-1.15}$ & $3.5^{+0.5}_{-0.5}$ &  14.2 & 0.25\\[0.4em]
WOCS 9005B (Best fit) & $11000^{+250}_{-250}$ & $0.063^{+0.001}_{-0.001}$ & $0.0525^{+0.0014}_{-0.0014}$ & $9.50^{+0.00}_{-0.25}$ & $14.28$ & 0.28 \\ [0.4em]
WOCS 9005B (Top 1 percentile) & 9000 -- 12000 & 0.036 -- 0.138 & 0.025 -- 0.113 & 6.5 -- 9.5 &  $<$ 30.94 & $<$ 21.5 \\ [0.4em] \midrule
\multicolumn{7}{c}{Kurucz\_UVBLUE + Koester} \\ [0.4em]
WOCS 9005A & $6500^{+500}_{-500}$ & $1.85^{+0.03}_{-0.03}$ & $5.52^{+1.15}_{-1.15}$ & $3.5^{+0.5}_{-0.5}$ & 14.1 &  0.17 \\[0.4em]
WOCS 9005B (Best fit) & $12500^{+250}_{-250}$ & $0.0221^{+0.0022}_{-0.0004}$ & $0.0108^{+0.0011}_{-0.0003}$ & $9.50^{+0.00}_{-0.25}$ & $28.03$ & 0.55 \\ [0.4em]
WOCS 9005B (Top 1 percentile) & 9750 -- 15250 & 0.009 -- 0.056 & 0.004-- 0.025 & 6.5 -- 9.5 &  $<$ 29.86 & $<$ 19.9 \\
\bottomrule
\end{tabular}

\flushleft
Note: RV$_{\text{COM}}=$ Center-of-mass radial velocity of binary, RV$_{semi}=$ Radial velocity semi-amplitude, e$=$ orbital eccentricity, a\,sini$=$ Projected semi-major axis, 
f(m) = ($M_{2}\,sini)^{3}$/($M_{1}+M_{2})^{2}$. $^{\dagger}$ Estimated through spectral analysis. $^{\ddagger}$ Obtained via PARAM, a web interface for the Bayesian estimation of stellar parameters (\href{http://stev.oapd.inaf.it/cgi-bin/param_1.3}{\url{http://stev.oapd.inaf.it/cgi-bin/param\_1.3}}) 
\end{table*}

\section{`Sighting' of WD in WOCS\,9005 from Spectral Energy Distribution} \label{sec:SED}

\begin{figure*}
    \centering
    \includegraphics[width =0.95\textwidth]{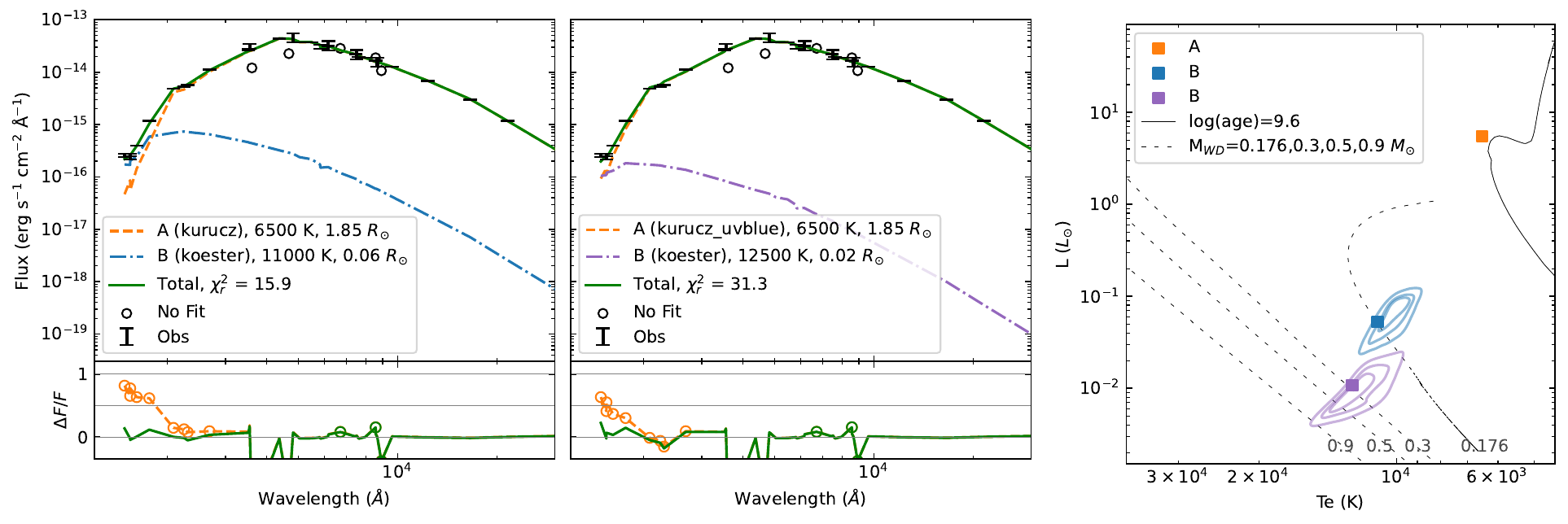}
    \caption{Binary SEDs of WOCS\,9005. The left panel shows the binary fit using Kurucz and Koester models. The middle panel shows the binary fit using Kurucz\_UVBLUE and Koester models. The fitted SEDs for A component (orange), B component (blue for Kurucz and purple for Kurucz\_UVBLUE), and total (green). The bottom sub-panel shows the fractional residual flux for the A component and total binary SED. The observed points are shown as black error-bars and data removed during fitting is shown as black circles.
    The right panel shows the H-R diagram of the components based on SED fit parameters, with same colors as the SEDs. The contours represent the top 1 percentile fits of the corresponding model. The PARSEC isochrone of log(age) 9.6 (gray curve) and WD cooling curves for WD masses of 0.176 \citep{Althaus}, 0.3, 0.5, and 0.9 \citep{Bergeron} are shown for reference.}
    \label{fig:SED}
\end{figure*}

An excess UV flux can indicate the presence of a hot companion star in a binary system, and an SED analysis can detect (1) the presence  of far-UV excess and (2) fit a double component SED to characterize the hotter and cooler binary companions, as shown in many recent studies \citep{subramaniam, uocsi, uocsii, uocsviii, Panthi2023MNRAS.525.1311P}. 
Following the technique used by \citet{uocsiv}, we constructed the SED of the source.
UVIT 
data are merged with observations from longer wavelengths using VOSA's Virtual observatory \citep{vosa}, to build a multi-wavelength SED.  
We obtained photometric fluxes in the range from UV to IR which include data from 
UVOT (UVW1, UVM2, UVW2, U) \citep{siegel3},  GALEX--GR6+7 (NUV) \citep{galex},\textit{HST} (F140LP, F150LP, F165LP; \citealt{nine2023}), APASS 9 (B,V,g,r,i) \citep{apass}, \textit{Gaia} DR3 (G, Gbp, Grp, Grvs; \citealt{vallenari2023gaia}), Pan Starrs PS1 DR2 (g, r, i, z, y) \citep{panstars}, SDSS DR12 (u, g, r, i, z), 2MASS All-Sky Point Source Catalog (J, H, Ks) \citep{2mass}, WISE (W1, W2, W3, W4) \citep{wise}. The fluxes are extinction (reddening) corrected based on the established extinction-wavelength relations given by \citet{fitzpatrick} and \citet{indebetouw}. We have adopted a distance modulus of 9.6 $\pm$ 0.04 mag, solar metallicity, and reddening E (B $-$ V) = 0.041 $\pm$ 0.004 mag \citep{Taylor}.

To estimate the parameters of WOCS\,9005, we first fitted the optically bright component (A) with two models using the \textsc{binary SED fitting} python code \citep{uocsiv}, which is based on $\chi^2$ minimization technique, i) Kurucz model \citep{caskurucz2003} and ii) a hybrid Kurucz\_UVBLUE model which uses UVBLUE model spectra \citep{Rodriguez2005ApJ...626..411R} within the 580--4700 \AA\ range and Kurucz spectra outside the wavelength range. Both models were fitted using parameter range as follows: log $g$ = 3--5, T$_{eff}$ = 3500--10000 K and [Fe/H] = 0 dex. Both models fitted the A component with similar parameters and the residual flux showed a significant UV excess as indicated by high fractional residuals ( $>50\%$), pointing to the `sighting' of the WD. Due to the substantial UV excess, we performed a double component fit for both the cases. We used the Koester model \citep{koester} to fit the hotter component (B), within a temperature range of 5000--80000 K and log $g$ range of 6.5--9.5.

The results of the binary SED fit are shown in Fig \ref{fig:SED}. and are tabulated in Table \ref{tab:abundances}. The T$_{eff}$, radius and the Log $g$ estimates of the cooler component are in excellent agreement with the spectroscopic estimates. We are able to `sight' and characterize the hot companion of WOCS\,9005 for the first time. 
The Kurucz and Kurucz\_UVBLUE model binary SEDs result in different parameters of the hotter component. The $\chi^2$ and fractional residual in Kurucz fitting is better compared to the Kurucz\_UVBLUE model.
Overall, the hot companion resembles a WD with T$_{eff}$ in the range of 11,000--12,500 K, with a radius of $\sim$ 0.02--0.06 R$_\odot$, and with a log $g$ of $\sim$ 9.5. The parameters of the cooler component  estimated using an SED-fitting technique by \citet[as reported in their table 1]{nine2023}, match well with our estimations. They did not detect UV excess for WOCS\,9005 based on the (F150N$-$F165LP, F150N) CMD, where the derived HST FUV narrow-band magnitude and color were compared with UVBLUE model based synthetic photometry. We used the same HST photometry in the observed bands, and detected noticeable excess UV flux ($\Delta F/F = 0.30$, 0.36 and 0.41 for F165LP, F150LP and F140LP, respectively) when compared to the UVBLUE model prediction (middle panel of Figure~\ref{fig:SED}). The discrepancy in identifying the UV excess in the source may be due to the different techniques (CMD vs SED) used to identify the excess and their usage of the derived narrow-bands.

The photometric mass was estimated by comparing the position of the hotter companion in the H-R diagram with WDs models \citep{Bergeron, Althaus}. The estimated mass of the companion WD is $\sim$ 0.176--0.3 $M_{\odot}$ and the log(age) $\sim$ 7.5 (for 0.3 $M_{\odot}$)--9 (for 0.176 $M_{\odot}$), based on the best fitted SED models. 
However, if we consider the top 1 percentile result of the $\chi^2$ fits, we obtain a wider range of WD parameters, resulting in temperature range of 9750--15250 K, mass range of 0.17-0.9 $M_{\odot}$ and log(age) range of 6.1--9. 
The kinematic mass of 0.5 $M_{\odot}$ falls within the above range and has corresponding cooling age of $\sim$60 Myr for the parameter range of Kurucz\_UVBLUE model fit. We therefore conclude that the SED analysis has indeed `sighted' the WD and characterized the hot companion. 
The WD cooling curves presented above assume that the WD has reached equilibrium and no more fusion is happening. However, as shown by \citet[Figure~4]{Althaus}, occasional flashes change the photometric parameters of the WDs. Thus, the above WD mass estimations depend on the evolutionary stage of the WD and unavailability of such information contributes to the uncertainty in the photometric mass estimation.
However, including both binary fits, it is found to a young system that has experienced a recent MT. Though the photometric mass may be less reliable, the youth of the system is confirmed. 

\section{Modeling \textit{(WOCS\,9005)}: A dwarf barium star} \label{sec:model}


MT in Ba {\scs II} stars from AGB companions predominantly occurs through stellar wind accretion rather than Roche lobe overflow \citep{boffin1988can}. Subsequent researchers have employed similar methods to develop both qualitative and quantitative models of the MT scenario in Ba {\scs II} stars \citep{cui2008orbital, husti2009barium, stancliffe2021formation}. In this section, we present the formation and evolution scenario of WOCS\,9005 through MT from an AGB star by adopting the wind accretion model of \citet{boffin1988can}.


WOCS\,9005 has a metallicity of [Fe/H] = -0.10 $\pm$ 0.05, suggesting that it likely had a solar-scaled abundance for each element before accretion. Therefore, the initial abundance of s-process elements should be [Ba/Fe]$_{initial}$ = 0,  [Y/Fe]$_{initial}$ = 0, and [La/Fe]$_{initial}$ = 0. 
After accretion, the observed abundances were [Ba/Fe] = 0.75 $\pm$ 0.08, [Y/Fe] = 1.09 $\pm$ 0.07, and [La/Fe] = 0.65 $\pm$ 0.06. Since the accreted matter is heavier than the initial envelope composition of the MS star, thermohaline mixing occurs, diluting the accreted matter into the entire envelope \citep[for a detailed explanation, see][]{stancliffe2007mass}. 
The initial envelope mass of approximately 3 times the mass transferred by the AGB ($M^T_{AGB}$; containing all the heavy metal) is derived for the accreting star using the formula given in eq. (1) of \citet{husti2009barium}. The present-day semi-major axis (A) based on the orbital parameters is 7.12 $\times$ 10$^{11}$ m. 
Although the semi-major axis can change following MT in the binary system, as suggested by \citet{boffin1988can} and \citet{husti2009barium}, treating a constant value of A is a reasonable assumption during the superwind phase experienced at the end of the AGB phase.
Based on these assumptions, the accreted mass calculated for a stellar wind velocity ($V_{\text{wind}}$) of 30 km s$^{-1}$ using the MT formula given in Eq. (6) of \citet{boffin1988can} \& Eq. (4) of \citet{husti2009barium} is approximately 0.15\,$\mathrm{M}_{\odot}$. The above value of $V_{\text{wind}}$ is consistent with the mean of the $V_{\text{wind}}$ in the range 20 to 40 km s$^{-1}$ estimated for the AGB stars \citep{knapp1985mass, pottasch1984}.
Consequently, based on the earlier relationship, the initial mass of the MS (now a Barium dwarf) star's envelope was $\approx$ 0.45 $\mathrm{M}_{\odot}$ and the final envelope mass after MT is $\approx$ 0.6 $\mathrm{M}_{\odot}$. 
This analysis serves our attempt to model WOCS\,9005 binary system using the method of \citet{boffin1988can}, where the barium dwarf would have formed via MT through stellar wind from AGB donor to MS star.
However, an in-depth analysis of WOCS\,9005 binary system in the future would offer greater insight into its characteristics, evolution, and MT mechanism.

\section{Discussion and conclusion} \label{sec:discussion}


There are several searches and characterization of dwarf barium stars using high resolution spectroscopic data \citep{roriz2024high, mardini2024metal}. 
There are very less number of known dwarf barium stars \citep{roriz2024high}. The T$_{eff}$ and the log $g$ range of the WOCS\,9005 falls in the range of these parameters of dwarf barium stars studied by \citet{roriz2024high}. WOCS\,9005 has relatively less abundances of Ba and Y in comparison to those in the three stars analyzed by \citet{roriz2024high}. 
The Y and La abundances of WOCS\,9005 are very similar to those in the giant barium star, NGC\,2420-173 ([Y/Fe] = 1.0; [La/Fe] = 0.62; P=1479 days; e=0.43 ; \citealt{jorissen2019barium}), which is found to be slightly metal poor ([Fe/H] = $-$0.26; \citealt{van2017mass}). 


Though our derivation of T$_{\rm eff}$ and log~$g$ for WOCS\,9005 matches excellently with \citet{nine2024wiyn} values, the barium enhancement of WOCS\,9005 was not uncovered by their analysis as they assume a higher value of $\xi_{t}$. The estimation of Ba abundance from the line EW/synthesis is sensitive to the assumed value of $\xi_{t}$ \citep{reddy2017}. We note that their estimated [Ba/Fe] values for WOCS\,8012 and 10014 (where they assume a lower value of $\xi_{t}$) are similar, whereas that for WOCS\,9005 differs by more than 0.6 dex with respect to  GALAH survey results.
 
 \citet{escorza2019barium} studied a number of field dwarf barium stars and their orbital properties. In their figure 9, the giant and dwarf barium stars are shown in the period-eccentricity diagram.  WOCS\,9005 with a period of 2769$\pm$18 days and eccentricity 0.15$\pm$0.05 falls in the densely populated region of known giant barium stars, though there are no known dwarf barium star in the region. The orbital parameters of NGC\,2420-173 fall within the similar range.
 Overall, WOCS\,9005 fits well with the orbital kinematics of the known barium stars. 

This is the first barium BSS star with a photometrically detected hot companion. This is also the first estimate of the WD properties and model based age of the WD companion of a barium BSS.


WOCS\,9005 and WOCS\,1020 are both BSSs with very similar properties, though WOCS\,1020 is a single member with FUV detection. Though many previous studies considered the above two stars as BSSs, \citet{geller2015} excluded them in their conservative list of BSSs. On the other hand, \citet{leiner2019} considered these two as blue lurkers, which are stars on the MS, but have experiences MT as inferred from their fast rotation. The panel (c) of Fig \ref{fig:galah}, confirms that these stars are indeed BSSs.
Among the MSTO comparison stars, WOCS\,4021, 5009 and 4004 are SB1s and have estimated orbital properties. Our analysis finds that these three SB1s on the MSTO have very similar abundances, that is also similar to the abundance of the BSS WOCS\,1020. 

We note that there is no detection of the three reference MSTO SB1s in the UVIT-FUV, even though they have periods in the range 4600 - 7400 days indicating that either the companions are much fainter MS stars or relatively older WDs. The detection of overabundance of s-process elements in WOCS\,9005 is due to a recent MT, as suggested by the young WD.
 We note that there are a few more SB1 systems with periods similar to WOCS\,9005 in M67 
but show a mild barium enhancement, as seen from figure 1, which we will explore in the future. 
It is possible that the abundance enhancement is a short lived phenomenon in BSSs.
The time-scale for diffusion or gravitational settling to effectively deplete the overabundance from the surface to the lower layers is likely to be faster for MS stars relative to giants due to their higher gravity, though there are no theoretical estimations. Missions such as UVIT/AstroSat, GALEX can only detect hotter and newly formed WDs. More sensitive missions such as the proposed INSIST mission \citep{subramaniam2022overview}, along with its spectroscopic capability, will be able to unearth WDs in many post-MT binary systems, including BSSs and barium stars.

\begin{acknowledgments}
    We thank the referee for valuable suggestions. AS acknowledges support from SERB power fellowship. VJ thanks the Alexander von Humboldt Foundation for their support.
    \textit{UVIT} project is a result of the collaboration between IIA, Bengaluru, IUCAA, Pune, TIFR, Mumbai, several centres of ISRO, and CSA. This publication makes use of {\sc VOSA}, developed under the Spanish Virtual Observatory project. 
    This work has made use of data from the European Space Agency (ESA) mission {\it Gaia} (\url{https://www.cosmos.esa.int/gaia}), processed by the {\it Gaia} Data Processing and Analysis Consortium (DPAC, \url{https://www.cosmos.esa.int/web/gaia/dpac/consortium}).
\end{acknowledgments}

%

\vspace{5mm}
\facilities{Astrosat(UVIT), Swift(UVOT), Gaia, AAT(HERMES), GALEX, AAVSO(APASS), PS1, Sloan, CTIO:2MASS, WISE}





\bibliography{sample631}{}
\bibliographystyle{aasjournal}






\begin{table*}
\centering
\caption{Elemental abundances [X/Fe] for 4 turn-off stars which were used to estimate avg abundances in Table \ref{tab:abundances}. The numbers in the parentheses are the number of lines employed for the abundance measurement of an element. The stellar parameters estimated using spectroscopy are shown below, followed by orbital parameters \citep{geller2021} and mass of both components.}
\label{tab:MSTO_abundances} 
\begin{tabular}{ccccc}  \toprule 
Parameter           &   WOCS 4021        &   WOCS 5009        &   WOCS 4004        &   WOCS 7010  \\  \midrule
$[$Na \scs{I}/Fe$]$ &  $+0.15\pm0.05(1)$ & $-0.02\pm0.05(1)$  & $+0.08\pm0.05(1)$  & $+0.13\pm0.05(1)$        \\
$[$Mg \scs{I}/Fe$]$ &  $+0.12\pm0.05(3)$ & $+0.10\pm0.04(2)$  & $-0.02\pm0.05(2)$  & $+0.01\pm0.05(3)$  \\
$[$Al \scs{I}/Fe$]$ &  $+0.08\pm0.03(3)$ & $+0.08\pm0.02(2)$  & $+0.05\pm0.05(4)$  & $+0.09\pm0.04(4)$  \\
$[$Si \scs{I}/Fe$]$ &  $-0.02\pm0.03(3)$ & $+0.04\pm0.02(4)$  & $-0.08\pm0.04(4)$  & $+0.09\pm0.04(4)$  \\
$[$Ca \scs{I}/Fe$]$ &  $+0.02\pm0.04(2)$ & $+0.01\pm0.05(3)$  & $+0.06\pm0.04(2)$  & $+0.05\pm0.06(3)$  \\
$[$Ti \scs{I}/Fe$]$ &  $-0.02\pm0.04(1)$ & $+0.01\pm0.05(3)$  & $-0.04\pm0.04(2)$  & $+0.05\pm0.05(3)$   \\
$[$Cr \scs{I}/Fe$]$ & $-0.04\pm0.03(1)$  & $-0.06\pm0.03(1)$  & $-0.02\pm0.03(1)$  & $+0.03\pm0.03(1)$   \\
$[$Fe \scs{I}/H$]$  & $-0.15\pm0.05(20)$ & $-0.15\pm0.05(21)$ & $-0.11\pm0.05(21)$ & $-0.16\pm0.05(20)$  \\
$[$Fe \scs{II}/H$]$ & $-0.16\pm0.06(2)$  & $-0.15\pm0.07(2)$  & $-0.11\pm0.07(2)$  & $-0.15\pm0.06(2)$   \\
$[$Ni \scs{I}/Fe$]$ & $-0.01\pm0.06(9)$  & $+0.07\pm0.06(7)$  & $-0.01\pm0.05(9)$  & $-0.05\pm0.06(7)$   \\
$[$Y \scs{II}/Fe$]$ & $+0.09\pm0.08(2)$  & $+0.07\pm0.08(2)$  & $+0.10\pm0.07(2)$  & $+0.08\pm0.07(2)$   \\
$[$Ba \scs{II}/Fe$]$& $+0.20\pm0.09(2)$  & $+0.15\pm0.09(2)$  & $+0.10\pm0.09(2)$  & $+0.12\pm0.09(2)$   \\
$[$La \scs{II}/Fe$]$& $+0.08\pm0.07(2)$  & $+0.09\pm0.07(2)$  & $-0.07\pm0.06(2)$  & $+0.08\pm0.06(2)$   \\ 
\midrule
T$_{\text{eff}}$ (K)$^{\dagger}$    & $6050 \pm 50$ & $5975 \pm 50$  & $6000 \pm 50$ & $6100 \pm 50$  \\
log~$g$ (cm s$^{-2}$)$^{\dagger}$   & $4.0 \pm 0.2$ & $3.9 \pm 0.2$  & $4.2 \pm 0.2$ & $4.0 \pm 0.2$  \\
$\xi_{\text{t}}$ (km s$^{-1}$)$^{\dagger}$ & $1.35 \pm 0.1$ & $1.24 \pm 0.1$ & $1.42 \pm 0.1$ & $1.66 \pm 0.1$ \\
$[$Fe/H$]$ (dex)$^{\dagger}$        & $-0.15 \pm 0.04$ & $-0.15 \pm 0.04$ & $-0.11 \pm 0.04$ & $-0.16 \pm 0.03$ \\
$v \sin i$ (km s$^{-1}$)$^{\dagger}$ & $10.0 \pm 2.0$ & $8.0 \pm 2.0$ & $8.0 \pm 2.0$ & $9.0 \pm 2.0$ \\

$M\ (M_{\odot}$)$^{\ddagger}$ & $1.23 \pm 0.03$ & $1.14 \pm 0.02$ & $1.38 \pm 0.06$ & $1.23 \pm 0.04$ \\
$R\ (R_{\odot}$)$^{\ddagger}$ & $2.03 \pm 0.05$ & $1.83 \pm 0.06$ & $2.46 \pm 0.14$ & $1.97 \pm 0.05$ \\
\end{tabular}
\begin{tabular}{ccccccccc}   
\toprule
Name &  RV$_{\text{COM}}$ & Period & RV$_{semi}$ & e & $f(m)$  & a\,sini  & $\mathrm{M}_{1}$ & $\mathrm{M}_{WD}$  \\
  & (km s$^{-1}$)     & (days) &(km s$^{-1}$)   &   & ($10^{-2} \mathrm{M}_{\odot}$) &(10$^{11}$ m) & ($\mathrm{M}_{\odot}$)  & ($\mathrm{M}_{\odot}$)  \\ \midrule
WOCS 4021& $34.46 \pm 0.16$ & $6166 \pm 21$ & $7.0 \pm 4.0$ & $0.83 \pm 0.07$ & $4.0 \pm 7.0$ & $3.40 \pm 1.90$ & $1.23 \pm 0.03$ & 0.49 \\  
WOCS 5009& $34.33 \pm 0.07$ & $4600 \pm 110$ & $1.65 \pm 0.09$ & $0.21 \pm 0.04$ & $0.20 \pm 0.03$ & $1.02 \pm 0.06$ & $1.14 \pm 0.02$ & 0.15 \\  
WOCS 4004& $34.75 \pm 0.19$ & $7400 \pm 300$ & $3.1 \pm 0.9$ & $0.70 \pm 0.14$ & $0.8 \pm 0.8$ & $2.2 \pm 0.8$ & $1.38 \pm 0.63$ & 0.28 \\  
\bottomrule
\end{tabular}

\flushleft
Note: RV$_{\text{COM}}=$ Center-of-mass radial velocity of binary, RV$_{semi}=$ Radial velocity semi-amplitude, e$=$ orbital eccentricity, a\,sini$=$ Projected semi-major axis, 
f(m) = ($M_{2}\,sini)^{3}$/($M_{1}+M_{2})^{2}$. $^{\dagger}$Obtained through spectral analysis. $^{\ddagger}$ Obtained via PARAM.

\end{table*}




\end{document}